\documentclass[useAMS,usenatbib]{mn2e}
\usepackage{graphicx}
\title{Collisional Hardening of Compact Binaries in 
Globular Clusters}
\author[S.~Banerjee and P.~Ghosh]
{
S.~Banerjee$^1$ and  
P.~Ghosh$^1$\\
$^1$Tata Institute of Fundamental Research, Mumbai, India.
}

\date{Submitted:                                                             ,
Accepted:                                            }
\pagerange{\pageref{firstpage}--\pageref{lastpage}} \pubyear{2006}

\newcommand{\Rsun}{\ensuremath{R_{\odot}}}
\newcommand{\Msun}{\ensuremath{M_{\odot}}}
\newcommand{\tpxb}{\ensuremath{\tau_{PXB}}}
\newcommand{\eg}{{\it e.g.}}
\newcommand{\ie}{{\it i.e.}}
\newcommand{\viz}{{\it viz.}}

\begin{document}
\label{firstpage}
\maketitle

\begin{abstract}
We consider essential mechanisms for orbit-shrinkage or ``hardening'' 
of compact binaries in globular clusters to the point of 
Roche-lobe contact and X-ray emission phase, focussing on the 
process of collisional hardening due to encounters between binaries 
and single stars in the cluster core. The interplay between this 
kind of hardening and that due to emission of gravitational 
radiation produces a characteristic scaling of the orbit-shrinkage 
time with the single-star binary encounter rate $\gamma$ in the 
cluster which we introduce, clarify, and explore. We investigate 
possible effects of this scaling on populations of X-ray binaries 
in globular clusters within the framework of a simple ``toy''
scheme for describing the evolution of pre-X-ray binaries in
globular clusters. We find the expected qualitative trends 
sufficiently supported by data on X-ray binaries in galactic
globular clusters to encourage us toward a more quantitative study.
\end{abstract}


\begin{section}{Introduction.}

It is well-known that globular clusters contain far more than their fair 
share of compact X-ray binaries per unit stellar mass, compared to their 
host galaxies (\citealt{vh87}, \citealt{vl2004}). The 
enhancement factor is $\sim 100$ in the 
Milky Way and M31 (\citealt{vl2004}, \citealt{p.et.al2003}), and 
possibly much higher in elliptical galaxies, 
as recent \emph{Chandra} observations have suggested 
(\citealt{a.et.al2001}, \citealt{p.et.al2003}). 
The origin of this overabundance of close binaries has been realized 
for some thirty years now to be the dynamical formation of such 
binaries --- through tidal capture and/or exchange interactions --- which 
can proceed at a very significant rate in dense cores of globular 
clusters (henceforth GCs) because of the high stellar-encounter rates 
there (\citealt{ht85}, \citealt{hv83}, \citealt{h.et.al92}), but whose 
rate is negligible over the rest of the galaxy, where the stellar density 
is low by comparison. The GC X-ray binaries that we 
shall be mainly concerned with in this work are those which are 
powered by accretion onto compact stars. These can be either (a) low-mass 
X-ray binaries (henceforth LMXBs), containing neutron stars accreting 
from low-mass companions, or, (b) cataclysmic variables (henceforth CVs), 
containing white dwarfs accreting from low-mass companions. Accordingly,
we shall not explicitly consider here binaries which contain 
either (a) two ``normal'' solar-mass stars, one or both of which are 
coronally active, or, (b) recycled neutron stars operating as 
rotation-powered millisecond pulsars, with a white-dwarf or a low-mass 
normal companion, although such binaries can be low-luminosity X-ray 
sources. However, general considerations on the dynamical formation of 
close binaries do apply to these as well; indeed, the latter binaries are 
now widely accepted as evolutionary products of LMXBs (\citealt{heu91,heu92}).
 
In tidal-capture formation of a compact-star binary, the compact star
(neutron star/white dwarf) passing close to a normal star dissipates its 
kinetic energy significantly by creating tidal deformation in the latter
star, and so becomes bound to it. In the exchange process of formation, 
a compact star replaces one of the stars of an existing binary system of 
two normal stars during a dynamical encounter (\citealt{ht85}, 
\citealt{hv83}, \citealt{spz}).
After formation in such an encounter, the compact-star binary continues 
to undergo stellar encounters in the dense cores of GCs,
and it is on one particular effect of the continuing encounters that
we focus in this paper. In the mid-1970s, it was realized that a major 
effect of the binary-single star encounters would be to extract energy 
from a given binary, making it more tightly bound or \emph{harder}, and 
giving this energy to the motion of the single stars in the GC, thus 
``heating'' the cluster (\citealt{h75}, \citealt{spz}, \citealt{h.et.al92}). 
We can call this effect \emph{collisional hardening} of the compact-star 
binary, which makes the binary's orbit shrink at a rate higher than that 
which would obtain if it were not subject to the above stellar encounters, 
\ie, if it were not in a GC. The latter rate is believed to be 
determined by a combination of two processes, \viz, (a) emission of 
gravitational radiation and (b) magnetic braking. In the former process, 
a compact binary emits gravitational radiation and so loses energy and
angular momentum, which makes its orbit shrink \citep{di92}. 
The latter mechanism is envisaged as follows. The low-mass companion to 
the compact star has a significant magnetic field, and also 
has its rotation tidally coupled or ``locked'' to that of orbital
revolution. The companion drives a wind, which carries away angular 
momentum at an enhanced rate because the magnetic field enforces 
co-rotation of the wind out to a radius considerably larger than that
of the star, and this angular momentum ultimately comes from the 
orbit because of the above tidal locking, thus making the orbit 
shrink \citep{vz81}. We discuss in this paper the relative roles of the 
above mechanisms for binary hardening, particularly the role of 
collisional hardening \emph{vis-\'a-vis} that due to gravitational
radiation, indicate and clarify a scaling that naturally emerges from 
this interplay, and briefly suggest possible observational signatures 
of this scaling. 

In Sec.~2, we discuss the hardening of compact binaries by the three
mechanisms discussed above, bringing out the particular role of
collisional hardening. We show that a characteristic scaling of the
orbit-shrinkage time of such binaries with an essential GC parameter
emerges because of the interplay
between collisional hardening and that due to gravitational 
radiation. In Sec.~3, we explore possible observational signatures
of this scaling. We sketch a very simple ``toy'' scheme for
describing the evolution of compact X-ray binaries in GCs, and indicate
a possible signature of the above scaling within the bounds of
this toy model. We show that current data on X-ray binaries in GCs are
consistent with this signature. We discuss our results in Sec.~4,
exploring possible lines of future inquiry.   
 
\end{section}

\begin{section}{Hardening of Compact Binaries.}

We consider the shrinking of the orbital radius of a compact binary by
the three mechanisms introduced above. Consider gravitational radiation
first. The rate at which the radius $a$ of a binary decreases due to 
this process is given by (see, \eg, d'Inverno 1992 and references 
therein):
\begin{equation}
\dot a_{\rm GW}\equiv\alpha_{GW}a^{-3}, \qquad
\alpha_{GW}\approx -12.2Mm_Xm_c~\Rsun/{\rm Gyr}
\label{GW}
\end{equation}
In this equation, $m_X$ and $m_c$ are respectively the masses of the 
compact star and its companion in solar masses, $M\equiv m_X+m_c$ is
the total mass in the same units, and the orbital radius $a$ is 
expressed in units of solar radius. We shall use these units
throughout this paper. 

Now consider magnetic braking. The orbit shrinkage rate due to this 
process is given in the original Verbunt-Zwaan prescription 
\citep{vz81} as:
\begin{equation}
\dot a_{\rm MB}\equiv\alpha_{MB}a^{-4}, \quad
\alpha_{MB}\approx -190{M^2\over m_X}
\left(R_c\over a\right)^4~\Rsun/{\rm Gyr}
\label{MB}
\end{equation}
where $R_c$ is the radius of the companion. We give below further 
discussion on this mechanism.

Finally consider the rate of orbit shrinkage due to collisional 
hardening, which is given by \citep{sh79},
\begin{equation}
\dot a_{\rm C}\equiv\alpha_{C}\gamma a^2, \qquad
\alpha_{C}\approx -2.36\times 10^{-7}{m_{GC}^3\over m_Xm_c}
~\Rsun/{\rm Gyr}
\label{COLL}
\end{equation}
Here, $m_{GC}$ is the mass of the single stars in the GC which are
undergoing encounters with the binary. In this introductory work, we 
assume $m_{GC}$ to be a constant, representing a suitable average value
for a GC core, which we take to be $m_{GC}\approx 0.6\Msun$.

The parameter $\gamma$ is a measure of the encounter rate 
between a given binary and the background of single stars in the core 
of the GC: it is a crucial property of the GC for our purposes, so
that we shall use it constantly in this work. 
It scales as $\gamma\propto\rho_c
/v_c$ with the (average) core density $\rho_c$ of the GC, and the 
velocity dispersion $v_c$ of the stars in the core. Following the  
convention often employed in the GC literature 
(\citealt{v2002}, \citealt{ht85}),  
we can, in fact, \emph{define} this parameter as:
\begin{equation} 
\gamma \equiv {\rho_c\over v_c}.
\label{DEFGAM}
\end{equation}
Then the unit of $\gamma$ is $\approx 6.96\times10^5 M_\odot 
R_\odot^{-4}$ sec, corresponding to the units of $\rho_c$ and $v_c$ 
commonly used in the GC literature, namely, $\Msun{\rm pc}^{-3}$ 
and km sec$^{-1}$ respectively. In these units, values of $\gamma$
generally run in the range $\sim 10^3-10^6$ (see below). 

In this work, we take the mass of the compact star 
to be $m_X=1.4\Msun$ and that of the companion $m_c=0.6\Msun$, the 
latter corresponding to a typical average mass of stars in a GC
(see above). 
According to the mass-radius relation for low-mass main-sequence 
stars, the radius of a main-sequence companion will then be  
$R_c\approx 0.6\Rsun$. Furthermore, we consider only circular 
orbits in this work, returning to this point in Sec.~4.

The total rate of orbit shrinkage due to the combination of the 
above mechanisms is given by: 
\begin{equation}
\dot a=\dot a_{\rm GW}+\dot a_{\rm MB}+\dot a_{\rm C}
\label{TOT}
\end{equation}
We emphasize that the first two terms in Eq.~(\ref{TOT}) are always 
operational, irrespective of whether the binary is in a GC or not, and 
it is the relative effect of the third term, which represents the
effects of the encounters in a GC core, that we wish to study here.
The interplay between the first and the third term was investigated
in a pioneering study by \citet{sh79}, before the magnetic braking
mechanism was postulated \citep{vz81}. 
\begin{figure}
\includegraphics[width=9.0cm,angle=0]{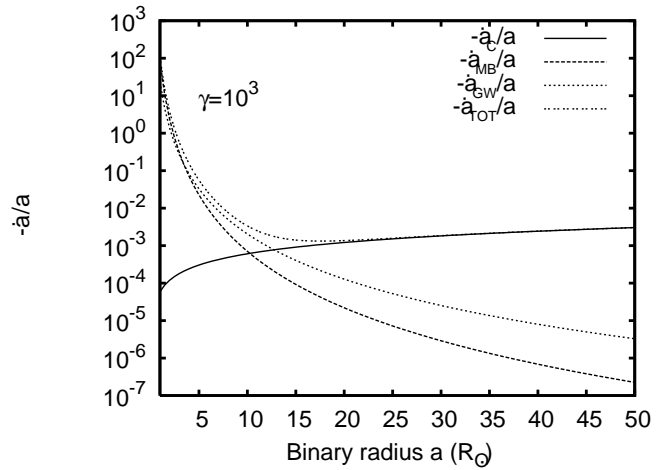}
\caption{\it Relative orbit shrinkage rates -$\dot a/a$ due to 
gravitational radiation, magnetic braking and collisional hardening, 
shown as functions of the binary separation $a$. Also shown is the 
total shrinkage rate. Value of $\gamma$ as indicated.}
\label{dominance}
\end{figure}

Note first that the three terms have different regions of dominance,
as shown in Fig.~\ref{dominance}. Collisional hardening dominates at 
large values of the orbital 
separation $a$, \ie, for wide binaries, while 
hardening by gravitational radiation and magnetic braking dominates 
at small $a$, \ie, for narrow binaries. Between the latter two,   
magnetic braking dominates at the smallest orbital separations, if
we adopt the original Verbunt-Zwaan (henceforth VZ) 
scaling for it (see below).
The relative orbit shrinkage rate $\dot a/a$ thus scales as $a$ at 
large orbit separations, passes through a minimum at a critical 
separation $a_c$ where the gravitational radiation shrinkage rate, 
scaling as $\dot a/a\sim a^{-4}$, takes over from collisional 
hardening, and finally rises at very small separations as    
$\dot a/a\sim a^{-5}$ due to VZ magnetic braking. The change-over
from gravitational radiation shrinkage to that due to magnetic 
braking occurs at a radius $a_m<<a_c$. These two critical radii 
are easily obtained from Eqs.~(\ref{GW}), (\ref{MB}), and 
(\ref{COLL}), and are given by  
\begin{equation}
a_c = \alpha_{\rm GW}^{1/5} \alpha_{\rm C}^{-1/5}\gamma^{-1/5},
\qquad a_m=\frac{\alpha_{\rm MB}}{\alpha_{\rm GW}}
\label{ACRIT}
\end{equation}
Note the scaling $a_c\propto\gamma^{-1/5}$, which is crucial for
much of our discussion here, as we shall see below. The critical 
orbital separation $a_c$ varies in the range $\sim(5-12)\Rsun$ for 
the canonical range of values of the above GC parameter 
\citep{v2002} $\gamma\sim 10^3-10^5$ in the above units 
\citep{sh79}.  
 
The relevance of this to close compact-star binaries in GCs is as
follows. When such a binary is formed, its orbital separation 
in most cases is such that the low-mass companion is not in 
Roche lobe contact, since the Roche-lobe radius has to be $R_L\sim
0.6\Rsun$ or less for this to happen for a typical low-mass main 
sequence or subgiant companion of mass $\sim 0.6\Msun$ (see above).
Mass transfer does not occur under such circumstances, so that 
such binaries are pre-LMXBs or pre-CVs, and we can call them by
the general name \emph{pre-X-ray binaries}, or PXBs for short. 
It is the above orbit-shrinkage or hardening process 
that brings the companion into Roche-lobe contact, so that
mass transfer begins, and the PXB turns on as an X-ray binary (LMXB
or CV), or XB for short.

Depending on the initial separation $a_i$ of the binary, some or all
of the above processes can thus play significant roles in shrinking 
it to the point where mass transfer begins. Recent numerical
simulations suggest that tidal-capture binaries are born with 
orbital radii (or semi-major axes) in the range $1<a_i/\Rsun<15$ for 
main-sequence (henceforth ms) or early subgiant companions, 
and in the range $40<a_i/\Rsun<100$ for horizontal-branch companions
\citep{z297}. For binaries formed by exchange encounters, 
the orbital radii are generally expected to be somewhat  
larger than those for corresponding tidal binaries with identical 
members. Thus, both collisional hardening and gravitational radiation 
are expected to play major roles in the orbital shrinkage to 
Roche-lobe contact for most of the PXBs in GCs, 
whether dynamically formed or primordial. 

Although we have included magnetic braking as above for completeness,
its role in hardening of PXBs into XBs appears to be rather 
insignificant, at least for the VZ scaling adopted above. This is
evident from the fact that there is little change in any of the 
results described here whether magnetic braking is included or not.
For the VZ scaling, this is easy to understand. With a steep increase 
at small $a$, this effect is significant only at very small orbital
separation, when the PXB has already come into Roche-lobe contact
and become an XB. Thus, this process may well be significant in the 
further orbital evolution of the XB as mass-transfer proceeds
(\citealt{heu91}, \citealt{heu92}), but not in the PXB-hardening 
process under study here.

Actually, further study of magnetic braking since the original VZ
formulation has revealed many interesting points. The nature and 
strength of this effect very likely depends on the mass and 
evolutionary state of the companion star. For example, magnetic
braking may become totally ineffective for very low-mass companions
with $m_c\sim 0.3\Msun$ or less, as these stars are fully 
convective. Whereas a significant convective envelope is necessary 
for strong magnetic braking, ``anchoring'' of the magnetic field in 
a radiative core is also believed to be essential for it, and it is
argued that the effect would basically vanish when the star becomes
fully convective (\citealt{sr83}, \citealt{ph02}). 
Indeed, this forms the basis for the standard explanation for the 
\emph{period gap} in CVs (\citealt{heu92} and references therein).
Further, studies of the rotation periods of stars in open clusters 
have suggested that magnetic braking may be less effective than 
that given by the VZ prescription: this has been modelled in recent
literature by either (a) the VZ scaling as above, but a smaller 
numerical constant than that given above, or, (b) a ``saturation''
effect below a critical value of $a$, wherein the scaling changes
from the VZ $\sim a^{-4}$ scaling of Eq.~(\ref{MB}) to a much slower
$\sim a^{-1}$ scaling below this 
critical $a$-value (\citealt{slu1}). While these 
modifications are of relevance to XB evolution, it does not appear
that they can alter the PXB-hardening results described here in any
significant way. Accordingly, we shall not discuss magnetic braking 
any further here, and keep this term in the complete equations only 
to remind ourselves that it is operational, in principle, for 
companions with $m_c\ge 0.3\Msun$.

\begin{subsection}{An Interesting Scaling}

Interplay between collisional hardening and gravitational-radiation
hardening near the above critical orbital separation $a_c$ produces 
a characteristic scaling, which we now describe. Consider 
the shrinkage time $\tpxb$ of a PXB from an initial orbital 
separation $a_i$ to the final separation $a_f$ corresponding to
Roche-lobe contact and the onset of mass transfer, given by: 
\begin{equation}
\tpxb(a_i,\gamma) \equiv \int_{a_i}^{a_f}
{da\over\dot a_{\rm GW}+\dot a_{\rm MB}+\dot a_{\rm C}}
\label{TPXB}  
\end{equation}
For given values of stellar masses, $\tpxb$ scales with the GC 
parameter $\gamma$ introduced above as 
\begin{equation}
\tpxb\sim\gamma^{-4/5}.
\label{SCALING} 
\end{equation}
The scaling is almost exact at high values of $\gamma$, \ie,  
$\gamma > 10^4$, say, there being a slight fall-off 
from this scaling at low $\gamma$'s.

\begin{figure}
\includegraphics[width=9.0cm,angle=0]{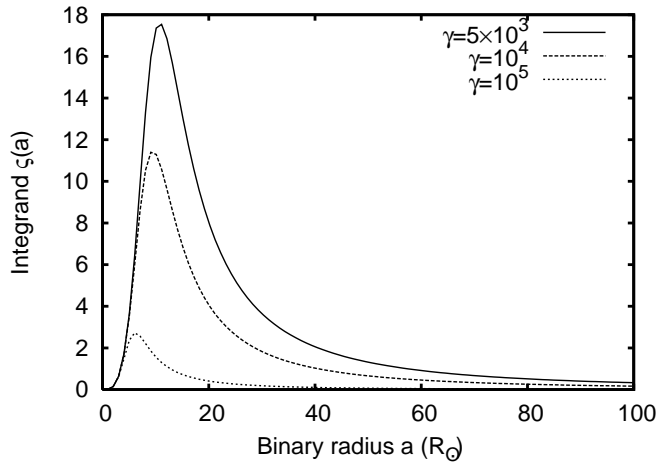}
\caption{\it Integrand $\zeta(a)$ in Eq.~{\rm (\ref{TPXB})} 
shown as function of orbital separation $a$, with values of 
$\gamma$ as indicated.}
\label{peak}
\end{figure}

How does this scaling arise? To see this, consider first the 
qualitative features of the integrand on the right-hand side of 
Eq.~(\ref{TPXB}), \ie, the reciprocal of the total shrinkage rate 
at an orbital separation $a$, which we denote by $\zeta(a)$, and 
which is displayed in Fig.~\ref{peak}. It is sharply peaked at 
$a\sim a_c$: indeed, the peak would be \emph{exactly} at the above 
critical separation $a_c$ but for the effects of magnetic braking, 
as can be readily verified. Since the latter effects are not 
important in the range of $a$-values relevant for this problem, 
as explained above, we can 
get a good estimate of the actual result by considering only 
gravitational radiation and collisional hardening. Because of this
dominant, sharp peak in $\zeta(a)$, most of the contribution to the 
integral, \ie, to $\tpxb$, comes from there, provided that 
the integration limits ($a_i,a_f$) are such that \emph{all or most 
of the peak is included}. We assume for the moment that this is so, 
and return to a discussion in the next subsection of what happens
when this condition fails.

Under the above circumstances, we can immediately give a 
rough estimate of $\tpxb$, which is the area under the curve in 
Fig.~\ref{peak}, as $\tpxb\sim 2a_c\times$(maximum value of the 
above integrand). This maximum value is simply $1/(2\alpha_{GW}
a^{-3})$ if we neglect magnetic braking, since the gravitational 
radiation term equals the collisional hardening term there, as 
explained above. This gives $\tpxb\sim\alpha_{GW}^{-1}a_c^4$, which, 
with the aid of Eq.~(\ref{ACRIT}), yields $\tpxb\sim\alpha_{GW}
^{-1/5}\alpha_C^{-4/5}\gamma^{-4/5}$. This is the basic reason for 
the scaling given by Eq.~(\ref{SCALING}).      

An exact evaluation of the integral in Eq.~(\ref{TPXB}), with the
magnetic braking term neglected, confirms this, as expected and as
detailed in Appendix A. The exact result is:
\begin{equation}
\tpxb = \alpha_{GW}^{-1/5}\alpha_C^{-4/5}\gamma^{-4/5}
[I(b_f)-I(b_i)].
\label{EXACT} 
\end{equation}
Here, $b$ is a dimensionless orbital separation defined 
by $b\equiv a/a_c$, and the integral
$I(x)$ is given in Appendix A. As $I(x)$ has only a logarithmic 
dependence on $x$ under these circumstances, the basic scaling is 
$\tpxb\sim\gamma^{-4/5}$, as above. It is this basic scaling that 
leads to the essential behavior of the shrinkage time $\tpxb$ 
discussed in this paper.

\end{subsection}

\begin{subsection}{Breakdown of Scaling?}

When would the above scaling break down, and why? A simple 
answer is clear from Fig.~\ref{peak}: this would happen when the 
integration limits ($a_i,a_f$) are such that all or most of the 
above peak in $\zeta(a)$ is \emph{not} included. For the present
problem, this basically reduces to an upper bound on $a_f$, since 
$a_i$ is normally large enough to ensure that the region of
integration in Fig.~\ref{peak} extends well into considerably 
larger values of $a$ beyond the peak.  
When $a_f$ becomes so large as to 
exceed $a_c$, the region of integration is severely curtailed 
from the left in Fig.~\ref{peak}, so that most of the peak's 
contribution is missed, and the above scaling breaks 
down. We might think that such a situation would arise 
when the low-mass companion in the PXB is an
evolved, horizontal-branch star, which has a much larger radius
than a ms/subgiant companion of the same mass, and so would be
expected to come into Roche-lobe contact at a much larger value 
of $R_L$, say $5-10\Rsun$, and correspondingly larger values of 
$a_f$. But such binaries are not relevant to our    
discussion here, since the lifetimes ($\approx 10^7$ y) of such 
horizontal-branch stars are too short to be of significance to 
the long binary-hardening timescales under consideration here. 
Thus, this possibility is not of practical importance here.

However, there \emph{is} a situation in which this scaling is not 
relevant, not because it breaks down, but, rather
because we move into a region of $\gamma$-values where $\tpxb$ 
computed in the above way exceeds the expected main-sequence 
lifetime $\tau_c$ of the low-mass ms/early-subgiant companion. 
Under these circumstances, the companion 
starts evolving into a giant and rapidly fills its Roche lobe, 
for essentially any value that $a$ is likely to have at 
that stage. This is formally equivalent to saying that $\tpxb$
saturates at a value $\tau_c$ in this range of $\gamma$. We
return to this point below.    
  
\end{subsection}

\begin{subsection}{Shrinkage Time}   
    
We now calculate the exact variation of the shrinkage time 
$\tpxb$ with the encounter-rate parameter $\gamma$ introduced 
earlier, keeping all terms in Eq.~(\ref{TPXB}). For this, we need 
to specify the initial and final values, $a_i$ and $a_f$, of the 
orbital separation. We adopt $a_f\approx 1.94R_{\odot}$ for 
ms/subgiant companions corresponding to Roche-lobe contact, when 
the radius of the Roche-lobe $R_L$ of the companion becomes 
equal to the radius of the companion itself. This translates into 
the above value of the orbital separation $a_f$ by the well-known 
Paczy\'nski relation:
\begin{equation}
R_L=0.46a\left({m_c\over M}\right)^{1/3}
\end{equation} 
corresponding to $R_L$ being equal to companion radius 
$\approx 0.6\Rsun$ for a companion mass $\approx 0.6\Msun$.
  
In general, $a_i$ will 
have a value which is within a possible range $(a_i^{min},
a_i^{max})$, which is indicated in Table \ref{tab1}. This range 
depends on the formation-mode of the binary, and also on the 
evolutionary status of the companion. The former has two 
possibilities, namely, (a) the binary is primordial, \ie, it 
was already a binary when the globular cluster formed, or, (b)
it formed by tidal capture or exchange interactions in the 
dense core of the globular cluster. The latter has also two basic 
possibilities, namely that the companion is either (a) a 
ms/subgiant, or, (b) a horizontal-branch star, as explained earlier.
As explained above, however, the short lifetimes of horizontal-branch 
(\citealt{k90}, \citealt{cl68}) stars compared to the timescales of 
hardening processes under study
here make it clear that they are of little importance in this 
problem, and we shall not consider them any further in this work. 
The ranges of $a_i$ adopted in various cases are detailed in Table 
\ref{tab1}, and are taken from current literature (\citealt{z297}).              
\begin{table}
\caption{\it Distribution functions $f(a_i)$ and range of 
initial orbital separations $a_i$ of compact-star binaries in  
globular clusters}
\begin{tabular}{lcl}
\hline
{\it Type of compact}&{\it Range of initial}&{\it Form of distribution}     \\
{\it binary}         &{\it radius} $a_i$    &{\it function} $f(a_i)$       \\
\hline
                     &                       &                                \\ 
Dynamically formed   &                       &   $f(a_i)\sim\frac{1}{a_i}$,   \\
compact star binary  &                       &   $f(a_i)={~\rm constant}$,    \\  
with ms or subgiant  & $2\Rsun$ - $50\Rsun$  &   $f(a_i)\sim a_i$,            \\
companion            &                       &   Gaussian in $a_i$            \\
                     &                       &   with $\mu=6.0\Rsun$ and      \\
                     &                       &   $\sigma=12.7\Rsun$           \\ 
                     &                       &                                \\
\hline 
                     
                     &                       &                                \\
Primordial compact   &$2\Rsun$ - $500\Rsun$  &  $f(a_i)\sim\frac{1}{a_i}$     \\
binaries             &                       &                                \\
                     &                       &                                \\
\hline
\end{tabular}
\label{tab1}
\end{table}
 
Since a GC has a distribution of $a_i$s, 
we wish to study how $\tpxb(\gamma,a_i)$ averaged over such 
a distribution scales with $\gamma$, since both of these represent 
overall properties of the cluster. To this end, we define a suitable 
average shrinkage time as:
\begin{equation}
\tau(\gamma)\equiv\langle\tpxb\rangle\equiv\int_{a_i^{min}}
^{a_i^{max}}\tpxb(\gamma,a_i)f(a_i)da_i,
\label{AVT} 
\end{equation}       
where $f(a_i)$ is the normalized distribution of $a_i$ in the range 
$(a_i^{min},a_i^{max})$. For this distribution, some indications and 
constraints are available, as follows.
For primordial binaries, the distribution $f(a)\propto 1/a$ 
(corresponding to a flat cumulative distribution in $\ln a$) is 
well-established \citep{k78}. For tidal capture/exchange 
binaries, results from numerical simulations like those of 
\citet{z297} generally suggest a bell-shaped distribution 
over the relevant ranges for both ms/subgiant companions and 
horizontal-branch companions, although other distributions are not
ruled out. To explore a plausible range of possibilities, we have
studied the following distributions, as detailed in Table \ref{tab1}: 
(a) the above reciprocal distribution  $f(a)\propto 1/a$, (b) a 
uniform distribution $f(a)=$ const, (c) a linear distribution $f(a)
\propto a$, and (d) a Gaussian distribution $f(a)\propto
\exp[-(a-a_0)^2/\sigma^2]$ with appropriately chosen central value 
$a_0$ and spread $\sigma$ given in Table~\ref{tab1}. 
The ultimate purpose is to assess the sensitivity (or lack thereof) 
of the final results on this distribution, as we shall see.
\begin{figure}
\includegraphics[width=9.0cm,angle=0]{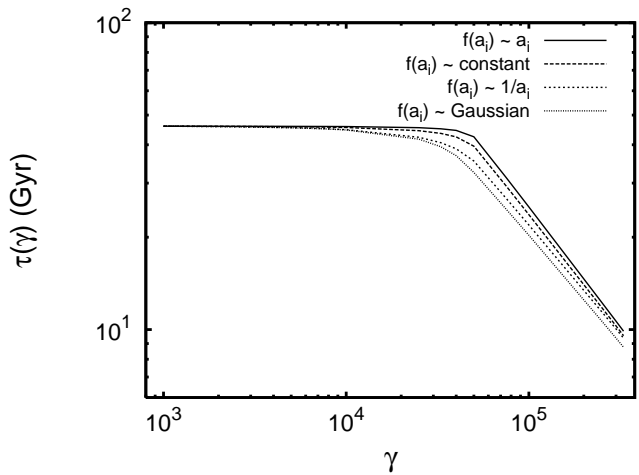}
\caption{\it $\tau(\gamma)$ vs. $\gamma$ for {\rm PXB}s: see text. 
Curves so normalized as to have the same ``saturation value'' 
$\tau_c=45{\rm~Gyr}$ at low values of $\gamma$.}
\label{tau}
\end{figure}

Calculation of $\tau(\gamma)$ clarifies the following points. 
Primordial binaries have a range of $a_i$s whose upper 
limit is considerably larger than that for ms/early-subgiant   
binaries, but most of those binaries which lie between these 
two upper limits are too wide to be of any practical importance 
in this problem. Thus, it appears that we need consider in detail
only PXBs with ms/early-subgiant companions for our purposes 
here, and Fig.~\ref{tau} shows the distribution-averaged shrinkage 
time $\tau(\gamma)$ as a function of the encounter-rate measure 
$\gamma$ for such binaries. As can be seen, the above 
$\gamma^{-4/5}$-scaling is almost exact at high values of
$\gamma$, say for $\gamma > 10^4$, there being a fall-off from 
this scaling at intermediate $\gamma$s, the extent of which 
depends on the case, as shown. We find that the above behavior 
can be well-represented by the analytic approximation
\begin{equation}
\tau(\gamma)\approx{A_0\over\gamma_0^{4/5} + \gamma^{4/5}},
\label{TAPPROX} 
\end{equation}
where $A_0$ is a constant which depends on the range $(a_i^{min},
a_i^{max})$ (and also on the stellar masses, as explained 
above), and $\gamma_0$ depends on the above and also on the 
distribution $f(a_i)$. To illustrate the latter effect, we have 
given in Table \ref{tab2} the inferred values of $\gamma_0$,
where the curves begin to deviate from the asymptote, for 
various distributions in the case of tidal capture/exchange 
binaries. 
 
We can see the trend that, as the distribution of $a_i$
tends to emphasize larger and larger values of $a_i$ over the 
permissible range (as happens in going from a $f(a_i)\sim 
a_i^{-1}$ to a uniform distribution $f(a_i)\sim$ const., and 
further to a linear distribution $f(a_i)\sim a_i$), 
$\gamma_0$ decreases. The physical reason for 
this is straightforward. Collisional hardening,
whose rate scales with $\gamma$, is dominant at large $a$ 
(scaling as $a^2$, as shown by Eqn.~(\ref{COLL})). Hence, larger
values of $a_i$ increase the relative contribution of collisional 
hardening to $\tau$, making it dominant over a larger range of
$\gamma$, so that the asymptote $\tau\approx A_0\gamma^{-4/5}$
corresponding to pure collisional hardening is followed over a
larger range of $\gamma$, and so $\gamma_0$ becomes smaller. It 
follows that those distributions which emphasize larger
values of $a_i$ will lead to smaller values of $\gamma_0$.
 
\begin{table}
\caption{\it Values of $\gamma_0$ obtained by fitting equation
$(\ref{TAPPROX})$ to computed $\tau(\gamma)$ vs. $\gamma$ curves
in Fig.~\ref{tau}}
\begin{center}
\begin{tabular}{|l|c|}
\hline
Type of initial               &  Value of          \\
distribution function         &  $\gamma_0$        \\
\hline
$f(a_i)\sim a_i$              &  $8.49\times10^3$  \\
\hline
$f(a_i)\sim {\rm constant}$   &  $8.74\times10^3$  \\
\hline
$f(a_i)\sim 1/a_i$            &  $1.21\times10^4$  \\
\hline
$f(a_i)\sim {\rm Gaussian}$   &  $1.06\times10^4$  \\
\hline 
\end{tabular}
\end{center}
\label{tab2}
\end{table}

Finally, at low values of $\gamma$ ( about $10^3$ - $3\times10^4$),
the following aspect of the low-mass companion's evolutionary 
characteristics enters the picture. The value of
$\tau(\gamma)$ calculated in the above manner then exceeds the
main-sequence lifetime $\tau_c$ of the companion, a simple,
widely-used estimate for which is 
\begin{equation}
\tau_c\approx 13\times 10^9\left({m_c/\Msun}\right)
^{-2.5}~{\rm yr}. 
\label{COMPLIFE} 
\end{equation}  
For a typical low-mass companion with $M_c\approx 0.6\Msun$,
therefore, $\tau_c\approx 45$ Gyr. When $\tau(\gamma)$ 
calculated as above exceeds this value of $\tau_c$, what 
happens is that the companion evolves into a giant,
and so comes into Roche-lobe contact at a time $\approx\tau_c$
for essentially all plausible values of $a$ at this point,
irrespective of the calculated value of $\tau(\gamma)$. 
This is formally   
equivalent to the statement that $\tau(\gamma)$ reaches a 
saturation value of $\tau_c$ at low values of $\gamma$, the 
change-over occurring at $\gamma = \gamma_c$ such that
$\tau(\gamma_c) = \tau_c$. Thus, the computed values 
can be analytically approximated by the prescription that 
$\tau(\gamma)$ is given by Eq.~(\ref{TAPPROX}) for $\gamma
>\gamma_c$, and by $\tau(\gamma) = \tau_c$ for $\gamma
<\gamma_c$. This is shown in Fig.~\ref{tau}, where we normalize 
all the curves to the above, common ``saturation value'' 
$\tau_c=45{\rm~Gyr}$. Note that the lifetimes 
of GCs are typically $\sim 10-14$ Gyrs, so that, in a 
given GC, only those PXBs which reach Roche-lobe   
contact within its lifetime would be relevant for our purposes.
What we have shown in Fig.~\ref{tau} is the \emph{formal} 
behavior of the distribution-averaged $\tau(\gamma)$ for 
plausible distributions of $a_i$. For a given GC, only that range of 
values of $a_i$ which corresponds to Roche-lobe contact within its
lifetime will go into the specific calculation for it.
   
\end{subsection}

\end{section}

\begin{section}{Evolution of Compact-Star Binaries in Globular 
Clusters}

How can we test the above scaling? Since $\tau$ is not directly
observable, are there possible signatures that its scaling with
$\gamma$ might leave in the observed behavior of the populations
of compact X-ray binaries in globular clusters? We briefly 
consider this question now and suggest possible answers.  

As remarked earlier, PXBs are formed in GCs primarily by tidal 
capture and exchange interactions in the GC core. 
The rate of the former process is proportional to 
the encounter rate between single stars in the GC core. The 
latter rate is commonly denoted by $\Gamma$ in the literature,
and it scales as $\Gamma\propto\rho_c^2r_c^3/v_c$ with the 
average core density $\rho_c$, the velocity dispersion $v_c$ 
of the stars in the core, and the core radius $r_c$. 
We can describe this as a rate of increase of the number $n_{PXB}$ 
of PXBs in the GC which is $\alpha_1\Gamma$, where $\alpha_1$ is a 
constant. In an exchange interaction, one of the members of a binary 
consisting of two normal stars is replaced by a heavier compact 
star. The rate of this process is proportional 
to the encounter rate between the above two populations. Assuming 
the population of compact stars in a GC to scale with the entire 
stellar population in the GC, and also the population of normal-star 
binaries in a GC to scale with its total population, both of which 
are normally done, the rate of the exchange process 
also scales with the square of the stellar density, and therefore
with the above $\Gamma$ parameter, and we can express it in a 
similar vein as $\alpha_2\Gamma$, where $\alpha_2$ is another 
constant. Thus, we can 
write the entire formation rate phenomenologically as 
$\alpha\Gamma$, where $\alpha\equiv\alpha_1+\alpha_2$.   

After formation, two main processes affect the fate of the PXBs.
The first is the process of the hardening of the PXB to the point 
of Roche-lobe overflow and conversion into an XB, as discussed 
in this paper. This proceeds on a timescale $\tpxb$, which means 
that the PXB population decreases on a 
timescale $\tpxb$, which we can describe phenomenologically by a 
rate of decrease of $n_{PXB}$ which is $n_{PXB}/\tpxb$. This 
process may be slightly modified by second-order ones, which are
normally ignored. For example, during the above hardening,  
an exchange encounter of the PXB with a single normal GC star 
heavier than the companion in the PXB can replace the latter. 

The second process is the destruction of a 
PXB by its encounter with single stars in the GC core. This can
happen in the two following ways. First, a fraction of the 
star-PXB encounters leads to a disruption or \emph{ionization} of 
the PXB. This leads to a reduction in $n_{PXB}$, whose rate is 
proportional to $\gamma$, the star-binary encounter rate introduced 
earlier, and also to $n_{PXB}$. We can thus express this rate of 
decrease in $n_{PXB}$ phenomenologically as 
$\beta_1\gamma n_{PXB}$, where $\beta_1$ is a constant. Secondly,
in a smaller fraction of such encounters, a compact star can 
replace the  normal low-mass companion in an exchange encounter, 
resulting in the formation of a double compact-star binary. This 
is equivalent to destroying the PXB, since such a binary will not 
evolve into an XB. The rate of this process scales with both 
$n_{PXB}$ and the rate of encounter between a given PXB and 
compact stars. If we again argue that the population of compact 
stars in a GC scales with the entire stellar population in it, 
this rate is $\propto\gamma$, and the rate of reduction of 
$n_{PXB}$ can be written phenomenologically as $\beta_2\gamma 
n_{PXB}$, where $\beta_2$ is a constant. The total PXB destruction rate 
can thus be written as $\beta\gamma n_{PXB}$, with 
$\beta\equiv\beta_1+\beta_2$.    

\begin{subsection}{A Simple ``Toy'' Evolutionary Scheme}  

We now combine the above points into a simple ``toy'' description
of PXB and XB evolution in GCs, which we can use in an attempt 
to extract possible signatures of the scaling described in this
work. In this ``toy'' scheme, which is similar in spirit to that 
of \citet{wh98} and \citet{gh2001} for following the evolution of 
X-ray binary populations of galaxies outside GCs, the evolution of 
the PXB population is given by:
\begin{equation}
{\partial n_{PXB}\over\partial t} = \alpha\Gamma - 
\beta\gamma n_{PXB} - {n_{PXB}\over\tpxb}.
\label{PXBEVOL} 
\end{equation}
wherein the above rates of increase and decrease of $n_{PXB}$ have 
simply been collected together.

The evolution of the XB population $n_{XB}$ resulting from the 
above PXBs is described in a similar manner: 
\begin{equation}
{\partial n_{XB}\over\partial t} = {n_{PXB}\over\tpxb} -
{n_{XB}\over\tau_{XB}}.
\label{XBEVOL} 
\end{equation}    
Here, $\tau_{XB}$ is the evolutionary timescale for XBs. The 
idea here is that XBs are created from PXBs at the
rate ${n_{PXB}/\tpxb}$, and conclude their mass-transfer phase,
and so their lifetime as XBs, on a timescale $\tau_{XB}$. 

In the spirit of the ``toy'' model, all timescales in equations 
(\ref{PXBEVOL}) and (\ref{XBEVOL}) can be considered constants, 
as can $\alpha$ and $\beta$, while in reality they depend on
orbital parameters and stellar properties, as also on other
parameters. These equations can then be solved readily, and of
interest to us here is the asymptotic behavior, obtained by
setting the time-derivatives to zero in these, which yields an
XB population:
\begin{equation}
n_{XB} = {\alpha\Gamma\tau_{XB}\over 1+\beta\gamma\tau(\gamma)}.
\label{XBASYM} 
\end{equation} 
The effect of collisional hardening is immediately seen on the 
right-hand side of Eqn.~(\ref{XBASYM}), in the second term in
the denominator. Since collisional hardening always decreases
$\tau$, it increases $n_{XB}$, other things being equal. This
enhancement in $n_{XB}$ is as expected, as collisional 
hardening makes a larger number PXBs reach Roche-lobe contact.
Thus, in a GC with given properties, the number of X-ray 
sources is expected to be enhanced by collisional hardening 
compared to what it would be if this effect were negligible.      

\end{subsection}

\begin{subsection}{Signature of Collisional Hardening?}

Can we look for observational evidence of the above enhancement 
in XB populations of GCs expected from collisional hardening? 
We discuss briefly an attempt to use \emph{Chandra} 
observations of GCs to this end, with the cautionary remark 
that our evolutionary model, as given in the previous subsection,
is still too simple-minded to apply quantitatively to actual GC
data. What we are looking for, therefore, is a possible 
\emph{qualitative} trend that is consistent with the ideas of 
collisional hardening discussed in this paper, which will
encourage us to perform a more detailed study.      

The trend given by Eqn.~(\ref{XBEVOL}) readily translates into one 
of the form 
\begin{equation}
{\Gamma\over n_{XB}} = A + B\gamma\tau(\gamma),
\label{XBASYM1} 
\end{equation}
where $A\equiv 1/\alpha\tau_{XB}$ and $B\equiv\beta A$ are 
constants, independent of the cluster parameters. We can 
compare this with data obtained from \emph{Chandra} observations 
of GCs, as given in \citet{p.et.al2003}. This is shown in   
Fig.~\ref{obs}.
 
\begin{figure}
\includegraphics[width=9.0cm,angle=0]{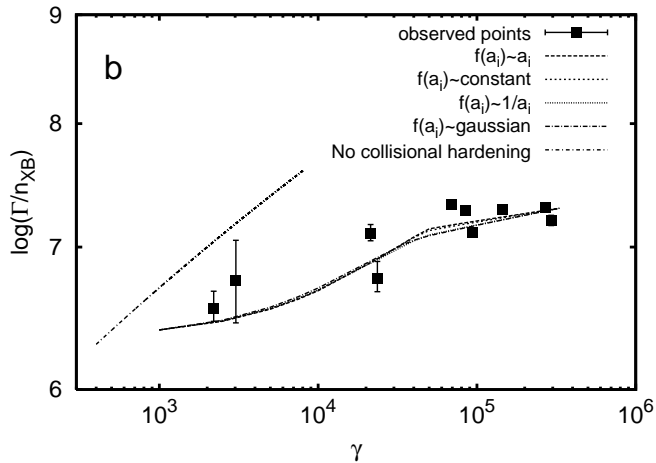}
\caption{\it $\Gamma/n_{XB}$ vs. $\gamma$ for galactic globular 
clusters. Observational points with error bars from 
\citet{p.et.al2003}. Trend suggested by Eq.~(\ref{XBASYM1}) shown
for various distributions as indicated. Also shown is the trend 
expected in absence of collisional hardening.}
\label{obs}
\end{figure}

It is clear that the trend suggested by Eq.~(\ref{XBASYM1}) is
consistent with the data, while that expected when collisional
hardening is entirely neglected is not. The flattening of the
trend on inclusion of collisional hardening is precisely due 
to the scaling discussed in this paper.
However, we stress again
that ours is only a ``toy'' model at this stage, relevant only
for exploring feasibility. To study the effect of $a_i$ 
distributions, we have normalized the constants $A$ and $B$ for
each distribution by having the curve pass through two chosen  
points at the lowest and highest values of $\gamma$ for which 
data is available. The results show clearly that varying the 
distribution has almost no effect on the trend.

\end{subsection}
      
\end{section}

\begin{section}{Discussions}

The purpose of this paper is to point out an essential effect in
the hardening of PXBs in GCs, \viz, that collisional hardening
\emph{increases} with increasing $a$ and orbital period, while 
that due to gravitational radiation has the opposite trend, so
that their interplay leads to a characteristic scaling of the 
total hardening rate with GC parameters. In our
introductory treatment of this effect here, we have given very
simple formulations of many physical processes. First, collisional 
hardening is an inherently stochastic process, wherein individual
events of varying sizes accumulate to yield a final state, and
the Shull (1979) rate we have used is a continuous approximation
to it. Secondly, the essential two- and three-body interactions 
that determine the evolution and fate of a PXB in a GC are also
stochastic by nature. For example, approximating an ionization 
event --- in which a binary is disrupted --- by a continuous 
term is necessarily a great oversimplification. Thus, an improved
treatment must include a proper formulation of these stochastic  
processes, which we are developing currently. 

Thirdly, we have confined ourselves to circular orbits here, 
while binaries created by tidal capture and/or exchange 
interactions often have quite eccentric orbits, in which tidal
circularization must play a dominant role during initial phases
of hardening. Fourthly, mass segregation is an essential effect   
in GCs, which reflects itself in the accumulation of the 
heaviest objects in the core of a GC, and so in a change in the 
effective mass-function in the core. Finally, the evolution of
the GC must be taken into account in a realistic calculation:
this would make the GC parameters time-dependent, while we have 
treated them as constants here, and may indeed have a large 
effect if core collapse and bounce are modelled.    

The above processes will need to be adequately modelled in a 
realistic calculation of the evolution of PXBs and XBs in GCs.
Such calculations are under way, and the results will be 
reported elsewhere.

\vskip6pt
\noindent
{\bf Acknowledgments:} It is a pleasure to thank the referee for 
thoughtful comments, which improved the paper considerably. 

\end{section}
%
%
%
%
 
%
%
%
\appendix
\section{Analytical expression for $\tau(\gamma)$}
We drop the magnetic braking term in the integral on right-hand side of 
Eqn.~(\ref{TPXB}), as explained in the text, and obtain:
\begin{equation}
\tau_{PXB}(a_i,\gamma) \approx \int_{a_f}^{a_i} \frac{da}
{\alpha_{\rm GW}a^{-3} + \alpha_{\rm C}a^2\gamma}=\frac{1}
{\alpha_{\rm GW}}I_1,  
\label{TPXB1}
\end{equation}
where,
\begin{equation}
I_1\equiv\int_{a_f}^{a_i}\frac{da}{a^{-3} + Ba^2\gamma},\qquad\qquad
B\equiv{\alpha_{\rm C}\over\alpha_{\rm GW}} 
\label{INT1}
\end{equation} 
Defining $\delta\equiv B\gamma$ and substituting $\delta a^5\equiv b^5$ in the 
above, we get 
\begin{equation}
I_1=\delta^{-{4\over5}}[I]_{b_f}^{b_i},		
\label{INT2}
\end{equation}
where the indefinite integral $I(x)$ is given by,
\begin{equation}
I=\int\frac{x^{3}dx}{1+x^{5}}
\label{INT3}
\end{equation}
Standard expressions for integrals of type $I$ are given in Gradshteyn and
Ryzhik (1980), from which we get,
$$I(x)=-{1\over5}\ln(1+x)
-{1\over5}\left[\cos{\pi\over5}\ln\left(1-2x\cos{\pi\over 5}+x^2\right)\right.$$
\begin{equation}
\left.+\cos{2\pi\over5}\ln\left(1+2x\cos{2\pi\over 5}+x^2\right)\right]
\label{INT4}
\end{equation}
$$+{2\over5}\left[\sin{\pi\over 5}\tan^{-1}\left(\frac{x-\cos{\pi\over 5}}{\sin{\pi\over5}}\right)
+\sin{2\pi\over 5}\tan^{-1}\left(\frac{x+\cos{2\pi\over 5}}
{\sin{2\pi\over5}}\right)\right]$$
From equations (\ref{TPXB1}), (\ref{INT1}), (\ref{INT2}), (\ref{INT4}),
\begin{equation}
\tpxb(a_i,\gamma)=\alpha_{\rm C}^{-{4\over5}}\alpha_{\rm GW}^{-{1\over5}}
\gamma^{-{4\over5}}\left[I\right]_{x=b_f}^{x=b_i}
\label{TPXB2}
\end{equation}
%
\label{lastpage}
\end{document}